\documentclass[12pt]{article}
\usepackage{times}
\usepackage{epsfig,graphicx}
\usepackage{amsmath}

\begin{document}
\title{Steady States of Epidemic Spreading in Small-World Networks}
\author{Xin-Jian Xu, Zhi-Xi Wu, Yong Chen, and Ying-Hai Wang\\
Institute of Theoretical Physics, Lanzhou University,\\
Lanzhou Gansu 730000, China}

\maketitle

\begin{abstract}
We consider a standard \textit{susceptible-infected-susceptible}
(SIS) model to study behaviors of steady states of epidemic
spreading in small-world networks. Using analytical methods and
large scale simulations, we recover the usual epidemic behavior
with a critical threshold $\lambda_{c}$ below which infectious
diseases die out. Whereas for the spreading rate $\lambda$ far
above $\lambda_{c}$, it was found that the density of infected
individuals $\rho$ as a function of $\lambda$ has the property
$\rho \approx F(K)(\ln \lambda - \ln \lambda_{c})$.
\end{abstract}

Complex networks have attracted an increasing interest recently.
The main reason is that they play an important role in the
understanding of complex behaviors in real world networks,\cite
{Strogatz, Barabasi_1, Dorogovtsev} including the structure of
language,\cite {Cancho, Sigman} scientific collaboration
networks,\cite {Newman_1} the Internet\cite {Albert, Huberman_1}
and World Wide Web,\cite {Huberman_2, Caldarelli} power
grids,\cite {Amaral} food webs,\cite {McCann, Williams} chemical
reaction networks,\cite {Alon} metabolic\cite {Jeong_1} and
protein networks,\cite {Jeong_2} etc. In the study of complex
networks, an analysis of the structures can give important
information about the underlying processes responsible for the
observed macroscopic behavior, such as small-world\cite {Watts_1}
and scale-free\cite {Barabasi_2} networks. In particular, social
networks have two characters. First, they show \lq\lq
clustering\rq\rq , meaning that two of your friends are far more
likely also to be friends of each other than two people chosen
from the population at random. Second, they exhibit the \lq\lq
small-world effect\rq\rq , namely, that any two people can
establish contact by going through only a short chain of
intermediate acquaintances. These two properties appear
contradictory because the first is a typical property of
low-dimensional lattices but not of random graphs or other
high-dimensional lattices, while the second is typically of random
graphs, but not of low-dimensional lattices. D.J. Watts and S.H.
Strogatz suggested a new small-world model\cite {Watts_1} recently
which interpolates between low-dimensional lattices and random
graphs and displays the both properties. In the model, a
small-world network is constructed as follows: starting with a
ring of $N$ vertices, each connected to its $2K$ nearest neighbors
by undirected edges, and then each local link is visited once with
the rewiring probability $p$ it is removed and reconnected to a
randomly chosen node. Duplicate edges are forbidden. After the
whole sweep of the entire network, a small-world network is
constructed with an average connectivity $\langle k \rangle = 2K$.
The WS networks has been widely studied because it constitutes an
interesting attempt to translate complex topologies of social,
economic, and physical networks into a simple model.

Although topological properties of complex networks have been
studied in detail,\cite {Barabasi_2, Newman_2, Barrat, Watts_2} a
natural question arises, that is the dynamical properties which
result from different networks. A good example is to inspect
complex features of epidemic spreading since the characterization
and understanding of epidemic dynamics in these networks could
probably provide us immediate applications to a large number of
problems, such as computer virus infections, distribution of
wealth, transmission of public opinion, etc. Recent papers\cite
{Kuperman, Agiza, Pastor, Newman_3} have given some valuable
insights of that: for small-world networks, there is a critical
threshold below which an infection with a spreading rate dies out;
on the contrary, for scale-free networks, even an infection with a
low spreading rate will prevalence the entire population.

In the present paper we consider a standard SIS model\cite
{Murray} in small-world networks, in which each node represents an
individual of the population and edges represent physical
interactions through which an infection spreads. According to the
SIS model, an individual is described by a single dynamical
variable adopting one of the two stages: \textit{susceptible} and
\textit{infected}. A susceptible individual at time $t-1$ will
pass to the infected state with the rate $\nu$ at time $t$ if it
is connected to one or more infected individuals. Infected
individuals at time $t-1$ will pass to the susceptible state with
the rate $\delta$ at time $t$, defining an effective spreading
rate $\lambda = \nu/\delta$. We can still keep generality by
setting $\delta = 1$. Individuals run stochastically through the
cycle susceptible $\rightarrow$ infected $\rightarrow$
susceptible, hence the model got its name. In the SIS model, an
important observable is the prevalence $\rho$, which is the time
average of the fraction of infected individuals reached after a
transient from the initial condition. Given a network, the only
parameter of the model is the spreading rate $\lambda$. The
information on the global spreading of the infection is contained
in the function $\rho(\lambda)$.

Small-world networks have very small diameters which mean the
presence of disordered long range interactions. In this case, the
networks are very homogeneous and quite reasonable that the
mean-field (MF) method is valid. By neglecting the density
correlations among the different nodes and ignoring all higher
order corrections in $\rho(t)$, the time evolution equation of the
SIS model can be written as\cite {Marro}
\begin{equation}
\dot{\rho}(t)=-\rho(t)+\lambda \langle k \rangle \rho(t)(1-
\rho(t)). \label{eq1}
\end{equation}
In the equation, the first term on the right-hand side (rhs)
considers infected nodes become healthy with unit rate and the
second term on the rhs represents the average density of newly
infected nodes generated by susceptible nodes. By imposing the
stationary condition $\partial_{t} \rho (t)=0$, one can obtain the
equation
\begin{equation}
\rho[-1 + \lambda \langle k \rangle (1-\rho)]=0 \label{eq2}
\end{equation}
for the steady state density of infected nodes $\rho$. The
equation defines an epidemic threshold
\begin{equation}
\lambda_{c} = \langle k\rangle^{-1}. \label{eq3}
\end{equation}
In other words, if the value of $\lambda$ is above the threshold,
$\lambda > \lambda_{c}$, the infection spreads and becomes endemic
with a finite stationary density $\rho$. Below it, $\lambda \leq
\lambda_{c}$, the infection dies out. In Euclidean lattices, J.
Marro and R. Dickman have concluded the order parameter behavior
of critical phenomena is $\rho \sim (\lambda-\lambda_{c})^{\beta}$
with $\beta \leq 1$ in the region $\lambda \sim \lambda_{c}$,\cite
{Marro} which uncovers the linear property in the critical
dimension. R. Pastor-Satorras and A. Vespignani also recovered
this property in small-world networks with the rewiring
probability $p=1.0$\cite {Pastor} recently; it is worth noticing
that in the extreme case the generated network is an entirely
random network with a restriction which leads to a large cluster.

\begin{figure}[h]
\centerline {\epsfxsize=9cm \epsffile{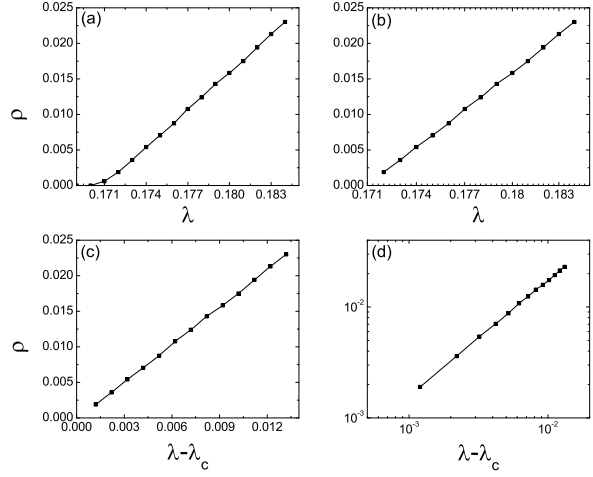}}

\caption{Density of infected nodes $\rho$ in the WS network with
$K=3$ from the simulations near by $\lambda_{c}=0.1708 \pm 0.004$,
which is in good agreement with the MF predictions $\lambda_{c}
=1/2K = 0.1667$. The numerical results were presented with $\rho$
vs $\lambda$ (a), $\rho$ vs $\log\lambda$ (b), $\rho$ vs $\lambda
-\lambda_{c}$ (c), and $\log \rho$ vs $\log(\lambda -\lambda_{c})$
(d). It indicates that all the plots perform the similar
behaviors.} \label{fig1}
\end{figure}

In order to compare with the analytical prediction we have
performed large scale simulations of the SIS model with parallel
updating in the WS network with the rewiring probability $p=0.1$.
The both properties of social networks are well presented by the
network in this case. The size of the network is $N=10^{6}$. The
number of the initially infected nodes is $10$ percent of the size
of the network. Simulations were implemented on the network
averaging over 20 different realizations. After an initial
transient regime, the systems stabilize in a steady state with a
constant average density of infected nodes. In Fig. \ref{fig1} we
plot the steady density of infected nodes $\rho$ with various
types of the scale of axis for $\lambda$ is very closed to
$\lambda_{c}$. The linear property of the order parameter of
Euclidean lattices, $\rho \sim (\lambda-\lambda_{c})^{\beta}$, is
well presented by the log-log plot of Fig. \ref{fig1}(d) which
gives the parameter value $\beta = 0.98 \pm 0.04$. However, it is
obvious that the four plots almost show the same shapes.
Consequently the following acceptable predictions for Fig.
$1$(a-d) are respectively given by
\begin{align}
\rho & \approx a_{1}\lambda +b_{1} \tag{4a}\label{eq4a}\\
\rho & \approx a_{2}\ln \lambda +b_{2} \tag{4b}\label{eq4b}\\
\rho & \approx a_{3}(\lambda -\lambda_{c})+b_{3} \tag{4c}\label{eq4c}\\
\ln\rho & \approx a_{4}\ln (\lambda -\lambda_{c})+b_{4}
\tag{4d}\label{eq4d}
\end{align}
Considering the Taylor expansion $\ln\lambda \approx
\frac{\lambda}{\lambda_{c}} +\ln\lambda_{c} -1$ at one order near
$\lambda_c$, Eq. (\ref{eq4b}) can be rewritten as
\begin{equation}
\rho \approx a'_{2} \lambda +b'_{2}. \tag{5}\label{eq5}
\end{equation}
One can also derive the equation
\begin{equation}
\rho \approx a'_{4}(\lambda - \lambda _{c}) + b'_{4}
\tag{6}\label{eq6}
\end{equation}
from the Eq. (\ref{eq4d}) in the same way. So, based on the
numerical simulations closed to $\lambda_{c}$, a analogous linear
relationship between the infected density and the spreading rate,
$\rho \approx a\lambda + b$, was extracted from the simple
approximation of series expansions (see Eq. (\ref{eq4a}), Eq.
(\ref{eq5}), Eq. (\ref{eq4c}), and Eq. (\ref{eq6}) corresponding
to Fig. \ref{fig1}(a)-(d), respectively). Therefore, we will
calculate the tendency of $\rho$ for $\lambda > \lambda_{c}$ next.

\begin{figure}[h]
\centerline{ \epsfxsize=9cm \epsffile{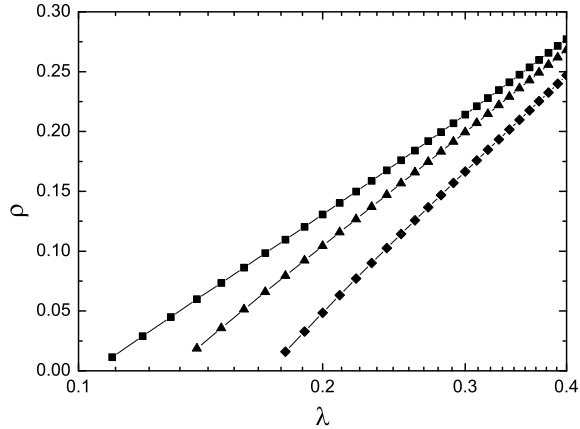}}

\caption{Extensive numerical results of the density of infected
nodes $\rho$ in the WS network for $\lambda
>\lambda_c$. All plots perform the identical behavior which is
described by Eq. (\ref{eq7}). Parameter values (from right to
left) $K=3$, $4$, $5$.} \label{fig2}
\end{figure}

\begin{figure}[h]
\centerline{ \epsfxsize=9cm \epsffile{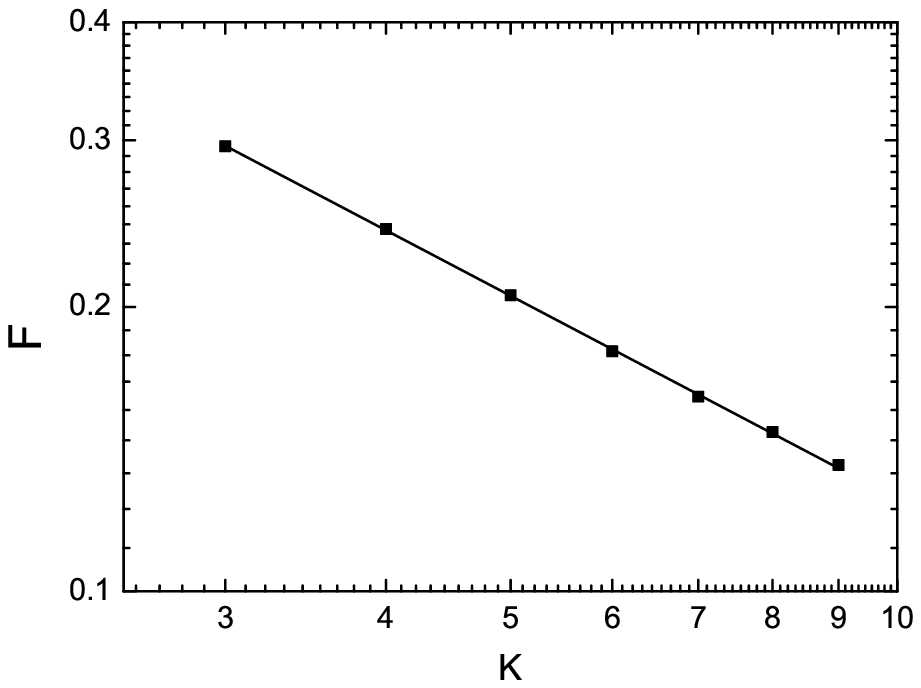}}

\caption{Log-log plot of the coefficient $F$ as a function of $K$.
The linear property of the plot predicts the power law, $F =
K^{-\alpha}$. Simulations give the parameter value $\alpha \approx
0.98$.} \label{fig3}
\end{figure}

As explicitly stated in Ref. \cite {Marro}, Eq. (\ref{eq1}) is
derived in the case of small values of $\rho$. So it is meaningful
only in the region $\lambda \sim \lambda_{c}$ and could not
determine the behavior for $\lambda$ far above $\lambda_{c}$,
i.e., the MF prediction of Euclidean lattices is invalid in this
region. In this condition, numerical methods are naturally adopted
to find the behavior of the prevalence for $\lambda >
\lambda_{c}$. In Fig. $\ref{fig2}$ we perform the numerical
results as far as $\lambda = 0.4$. The linear property of Eq.
(\ref{eq4d}) was presented excellent agreement with the numerical
results. Note that the constant parameters $a_{4}$ and $b_{4}$ in
Eq. (\ref{eq4d}) should be taken count of the initial setting
value of the average connectivity $2K$. More exact conclusion for
the nature of Fig. \ref{fig2}, which presented the whole behavior
at $\lambda
>\lambda_{c}$, should be given by
\begin{equation}
\rho \approx F(K) (\ln \lambda - \ln \lambda_{c}), \quad \text{if}
\quad \lambda
> \lambda_c. \tag{7}\label{eq7}
\end{equation}
To complete our study of the steady states of the SIS model in
small-world networks, we compute the coefficient $F(K)$ in Eq.
(\ref{eq7}). In Fig. \ref{fig3} we plot the coefficient $F(K)$ as
a function of $K$. The linear behavior of the log-log plot
predicts the power law, $F=K^{-\alpha}$. Numerical results give
the parameter value $\alpha \approx 0.98$.

In summary, we have analytically and numerically studied the
steady states of epidemic spreading in small-world networks. In
the region $\lambda \sim \lambda_{c}$, we recover the MF results
of Euclidean lattices, $\rho \sim (\lambda -
\lambda_{c})^{\beta}$. But when spreading rates become far above
the threshold, $\lambda
> \lambda_{c}$, the MF method can not work normally and the numerical method
is adopted. It was found that the behavior of order parameter has
the property $\rho \approx F(K)(\ln \lambda - \ln \lambda_{c})$.
In the present work, we just consider the dynamics on the WS
network with the rewiring probability $p=0.1$. However, for other
values of $p$, the results are qualitatively and quantitatively
the same as that we get.

We thank Prof. Hong Zhao for valuable discussions. This work was
partly supported by the National Natural Science Foundation of
China under Grant No. $10305005$ and the Natural Science
Foundation of Gansu Province.

\newpage

\bigskip

\end{document}